# Self-induced White-light Seeding Laser in A Femtosecond Laser Filament

Wei Chu[1,2], Guihua Li [2,3], Hongqiang Xie [2,3], Jielei Ni [2], Jinping Yao [2], Bin Zeng [2], Haisu Zhang[2,3], Chenrui Jing[2,3], Huailiang Xu[1,4], Ya Cheng[2,5], and Zhizhan Xu[2,6]

[1]State Key Laboratory on Integrated Optoelectronics, College of Electronic Science and Engineering, Jilin University, Changchun 130012, China

[2]State Key Laboratory of High Field Laser Physics, Shanghai Institute of Optics and Fine Mechanics, Chinese Academy of Sciences, P.O. Box 800-211, Shanghai 201800, China

[3]Graduate School of Chinese Academy of Sciences, Beijing 100039, China

[4] huailiang@jlu.edu.cn
[5] ya.cheng@siom.ac.cn
[6] zzxu@mail.shcnc.ac.cn

**Abstract:**

We report, for what we believe to be the first time, on the generation of remote self-seeding laser amplification by using only one 800 nm Ti:Sapphire femtosecond laser pulse. The laser pulse (∼40 fs) is first used to generate a filament either in pure nitrogen or in ambient air in which population inversion between the $B^2\Sigma_u^+$ and $X^2\Sigma_g^+$ states of $N_2^+$ is realized. Self-induced white light inside the filament covering the $B^2\Sigma_u^+$ - $X^2\Sigma_g^+$ transition is then serving as the seed to be amplified. The self-induced narrow-band laser at 428 nm has a pulse duration of ∼2.6 ps with perfect linear polarization property. This finding opens new possibilities for remote detection in the atmosphere.

1. Introduction

Femtosecond laser filamentation, which originates from a dynamic balance between Kerr self-focusing and the defocusing effect of the plasma generated at the self-focus location, can be managed to occur at a distance as far as a few kilometers in the atmosphere [1]. Furthermore, filaments can propagate without perturbation in adverse conditions such as rain [2], fog [3], turbulence [4, 5], which makes them highly suitable for atmospheric remote sensing [6, 7]. Currently identified remote targets based on high laser intensity inside filament in laboratory scale include all phases of samples such as $CH_4$ [8] and $C_2H_2$ [9] gases, ethanol vapor [10], bio-agents such as grain dusts of barley [11] and powders of egg white and yeast [12], water aerosols with multiple solutes (NaCl, $PbCl_2$, $CuCl_2$ and $FeCl_2$) [13, 14] and metallic targets [15, 16]. In the abovementioned remote sensing technology, characteristic fingerprint spectra from remote may be masked by the supercontinuum emission generated inevitably during the filamentation process, which spans from ultraviolet to infrared. This would lead to a lower signal-to-noise ratio of this technique in remote sensing. Recently, the observations of backward lasing actions by amplified spontaneous emission (ASE) during filamentation in air [17] or other gases [18] indicate a distinct promotion of signal intensity, and also a unique free-space 'incoherent' amplified light source, inspiring strong interest in the development of remote lasers for atmospheric applications.

More recently, by using infrared pump laser light at $\sim 1.9\,\mu m$, remote forward laser actions have been demonstrated [19, 20]. Different from the previously reported ASE schemes which result from the electron-ion recombination [17], thermal electron collisional excitation [21, 22], or resonant excitation transfer [18] etc., this forward laser scheme is based on stimulated amplification seeded by high-order harmonics generated during the filamentation of the infrared pump laser [23] with population inversion established in an ultrafast time scale comparable to the pump laser pulse [24, 25]. However, although the seed amplification scheme provides a new type of 'coherent' amplified light source in remote free-space without using any cavity optics, the limited output energy of the infrared laser light from optical parametric amplifier (OPA) makes it hard to generate a filament at a far distance.

Since current commercially available high-power TW laser systems for atmospheric studies are mainly Ti: sapphire laser systems with pulse duration of about tens of femtoseconds and the wavelength at ∼800 nm [26], it is naturally anticipated to directly generate abovementioned remotely coherent lasers with such Ti:sapphire laser systems. However, previous studies on remotely coherent lasers by using Ti:sapphire laser systems only focused on the proof-of-concept demonstration of the feasibility for realizing population inversion in a molecular ion system by using a pump-probe scheme with the probe pulse serving as the seed [24, 25]. Because it is, due to the temporal and spatial overlapping problems of two pulses, of particular difficulty to generate the seed amplification by using a two-beam scheme at a remote distance in a femtosecond laser filament, the present study is thus aiming at showing the possibility to produce coherent amplification light source in free space with only one femtosecond Ti: sapphire laser beam.

In this work, we demonstrate that coherent light amplification actions can indeed be induced in a plasma filament by using only one femtosecond Ti: sapphire laser pulse at 800 nm. We observe that the laser pulse, during its filamentation process either in pure nitrogen or in air, generates a plasma channel in which nitrogen ions at both ground state ($X^2\Sigma_g^+$) and excited state ($B^2\Sigma_u^+$) are prepared with population inversion. Meanwhile, through spectral broadening mechanisms such as self-phase modulation, the laser pulse transforms itself into a white light pulse whose spectrum can cover the transition of nitrogen ion between the $B^2\Sigma_u^+$ and $X^2\Sigma_g^+$ states, triggering stimulated amplification of spectral portion of white light at wavelengths corresponding to the $B^2\Sigma_u^+$ - $X^2\Sigma_g^+$ transition of nitrogen ion. The simple self-seeding laser scheme provides a further step towards the practical applications of remote lasers for stand-off spectroscopy of trace molecules in the atmosphere.

2. **Experimental setup**

The experimental schematic is shown in Fig. 1. A Ti:sapphire laser system consisting of an oscillator (Micra, Coherent) and an amplifier (Legend, Coherent) is used to produce the linearly polarized laser pulse with the center wavelength at 800 nm, a pulse duration of ~40 fs and a repetition rate of 1 kHz. The maximum energy of the output pulse is ~6 mJ, which can be controlled continuously by rotating two 500-μm-thick fused-silica windows inserted into the pump beam. The laser pulse is focused by a fused silica lens (L1) with a focal length of 50 cm into a gas chamber filled with pure nitrogen gas at 200 mbar, forming a filament. Note that the critical power is ~ 50 GW at this working pressure if assuming the value of the critical power in air to be 10 GW [27], which corresponds to an input laser energy of ~2.0 mJ when the laser pulse duration is ~40 fs. After the filament, the output light signal are collected and collimated by another fused silica lens (L2) with a focal length of 40 cm. A dichroic mirror with high reflectivity at 400 nm and high transitivity at 800 nm is then used to reflect the collimated beam in order to filter out the fundamental 800 nm laser light. The reflected signal by the dichroic mirror is detected by an imaging grating spectrometer (Shamrock 303i, Andor). In the measurements, the dichroic mirror can also be replaced with a gold mirror in order to record the full spectrum of the output laser pulse from the filament. For the experiment performed in air, all the components in the setup are kept at the almost same positions as before, except the gas chamber is removed and the fundamental 800 nm laser is focused in air directly.

3. **Results and discussion**

Fig. 2(a) demonstrates the full spectrum of the laser pulse output from the filament in a semi-logarithmic plot, in which the super-continuum (white light), and also a distinct and strong narrow-bandwidth emission at ~428 nm can be clearly observed. The input pump laser energy is ~5.5 mJ, and the spectrum is averaged over 500 laser shots. It can be seen in Fig. 2 (a) that the strong white light spectrum extends to the blue side down to 500 nm, with a long and very weak tail covering the wavelength of 428 nm. The broad spectrum of white light is also an evidence for the formation of filamentation in the nitrogen gas chamber. By finely tuning the distance between the two gratings in the compressor, the spectral intensity of the supercontinuum emission as well as the signal intensity of the 428 nm narrow-bandwidth emission can be optimized.

The narrow-band 428 nm emission corresponds to the P branch band head ($\Delta J = -1$) of the vibrational transition between the $B^2\Sigma_u^+(\upsilon = 0)$ and $X^2\Sigma_g^+(\upsilon = 1)$ states of $N_2^+$, as illustrated in Fig. 2 (b). In addition to the strong 428 nm emission, a weak band signal at the blue side of the 428 nm peak can be observed, as shown in Fig. 2(c), in which the emissions around 400 nm is separated from the strong red part of the supercontinuum spectrum by using a dichroic mirror with high reflectivity at the wavelength range from 350 nm to 450 nm and a QB-11 glass filter. The weak band signal corresponds to the R branch ($\Delta J = 1$) transition of the $B^2\Sigma_u^+(\upsilon = 0)$ and $X^2\Sigma_g^+(\upsilon = 1)$ states of $N_2^+$.

It can be seen in Fig. 2(a) that the peak intensity of the 428 nm band is comparable to that of the white light generated during the filamentation. This phenomenon is ascribed to seed amplification in the filament, similar to those reported previously in the pump-probe scheme by using two color laser beams [24, 25]. In the current study, the laser pulse first generates a plasma filament in the pure nitrogen gas, in which nitrogen ions at both ground state ($X^2\Sigma_g^+$) and excited states ($B^2\Sigma_u^+$) are prepared with the population inversion [25]. Meanwhile, during the filamentation, the laser pulse transforms itself into a white light pulse whose spectrum covers the transition between the $B^2\Sigma_u^+(\upsilon = 0)$ and $X^2\Sigma_g^+(\upsilon = 1)$ states, triggering stimulated amplification.

We call this self-induced white-light seeding amplification phenomenon in the filament as 'self-seeding laser' action.

To give the pattern of the self-seeding 428 nm laser beam, the isolated 428 nm laser signal is captured in a far field by a digital camera (Nikon, D40), as shown in Fig. 2(d). It can be clearly seen that the image of the 428 nm beam profile doesn't show a Gaussian profile, but a donut-like pattern. This strange donut-like pattern may result from the intensity distribution of the laser pulse inside the filament core.

In order to characterize the 428 nm laser line, we measure its polarization property by placing a Glan-Taylor polarizer just before the imaging spectrometer. It is found that the signal shows a nearly perfect linear polarization property with the polarization direction parallel to that of the pump pulse, as shown in Fig. 3 (a), in which zero degree indicates the polarization direction of the pump laser pulse. Since the white light generated during the filamentation has a polarization direction as that of the pump pulse, this result is similar to that obtained in our previous observation with the harmonic of the infrared light as the seed [19]. We also measured the 428 nm laser intensity as a function of the input laser energy, and it is found that the 428 nm laser signal is exponentially proportional to the pump laser intensity, as can be seen in Fig. 3 (b), showing that the laser intensity can be increased significantly for practical application if a high power laser with a reasonable laser energy, for example, 100 mJ, is employed.

The temporal property of the 428 nm laser is measured with a cross-correlation method by recording the sum frequency of the 428 nm laser pulse with a second 800 nm laser pulse, which is generated by using a 2-mm-thick BBO crystal. The experimental setup is shown in Fig. 4 (a). A small portion (5%) of the fundamental 800 nm beam is split from the output beam of the laser system with an energy of ~0.3 mJ, and used as a reference pulse in the cross-correlation measurement. The remaining portion of the 800 nm beam (~5.6 mJ) is used to generate the 428 nm laser line with the same scheme as described in Fig. 1. After passing through the glass filter, the 428 nm laser pulse, which is measured to be ~0.1 μJ, is combined with the 800 nm reference pulse by using another dichroic mirror. The combined signals are then sent into the 2-mm-thick BBO crystal for sum frequency generation, which is recorded using the imaging grating spectrometer.

Fig. 4(b) shows a typical sum frequency spectrum of the 800 nm reference pulse with the 428 nm laser pulse. By adjusting the delay between the 800 nm reference pulse and the 428 nm laser pulse, the sum frequency signal intensity changes, as shown in Fig. 4 (c), in which the zero delay is indicated as the maximum of the sum frequency intensity where complete overlap of the two pulses occurs; while positive delay represents that the 428 nm laser pulse is behind the 800 nm reference pulse. Based on the cross-correlation measurements, the full width at half maximum (FWHM) of the measured delay curve in Fig. 4(c) is determined to be ~2.6 ps. Since the pulse duration of the 800 nm reference pulse is ~40 fs which is much shorter than the measured value, the duration of the self-seeding 428 nm laser pulse can be approximated to the value of ~2.6 ps (FWHM). It should be pointed out that the measured pulse duration is much shorter than those generated based on ASE schemes whose pulse durations are generally in the range of a few ns [18].

In order to demonstrate the capability of the generation of such self-seeding laser in the atmosphere by using only one Ti:sapphire laser beam at 800 nm, we also carry out a measurement by directly focusing the fundamental 800 nm Ti:sapphire laser in air. In this case, a full spectrum of the laser pulse output from the filament in the range of 380-470 nm is recorded. As shown in Fig. 5, besides the strong white light, distinct laser signals at 391 nm and 428 nm at the blue tail of the supercontinuum spectrum can be clearly observed, which correspond to the vibrational transitions of $B^2\Sigma_u^+$ ($\upsilon=0$) - $X^2\Sigma_g^+$ ($\upsilon=0$) and $B^2\Sigma_u^+$ ($\upsilon=0$) - $X^2\Sigma_g^+$ ($\upsilon=1$), respectively. The multi-wavelengths laser lines results from the available seed signals from the supercontiuum spectrum in air, which extends up to 380 nm and is much broader than that observed in Fig. 2(a). As a consequence, this measurement clearly shows the feasibility for generating remote self-seeding nitrogen laser in air only by one femtosecond Ti:sapphire laser pulse.

4. Summary

In summary, we have demonstrated that, with only one 800 nm Ti: sapphire laser beam, ultrafast sub-10ps, self-seeding cavity-free nitrogen laser can be generated during the filamentation both in pure nitrogen and in air. It was found that the 800 nm pump laser plays two roles in the self-seeding laser scheme; one is to establish the population inverted ion system in the filament, and the other is to generate the supercontinuum seed light which can cover the population-inverted transition of molecular ions for subsequent amplification. Different from the previously reported atmospheric laser scheme based on harmonic seed amplification by using optical parametric laser source with limited output energy, the use of only one 800 nm Ti: sapphire laser beam to generate such remote self-seeding cavity-free laser in air provides a new possibility for remote detection and atmospheric applications.
.

**Acknowledgements**

This work is financially supported by National Basic Research Program of China (Grant 2011CB808102), National Natural Science Foundation of China (Grant Nos. 11134010, 11074098, 61235003, 60825406, 10974213, and 11204332), the Open Fund of the State Key Laboratory of High Field Laser Physics (SIOM) and the Fundamental Research Funds of Jilin University.

**Figure captions**

Figure 1. (Color online) Experimental setup for self-seeding laser generation.

Figure 2. (Color online) (a) Experimentally measured spectrum of super-continuum with the strong narrow-band emission at ~428 nm. (b) Energy-level diagram of ionized and neutral nitrogen molecules in which the vibrational transition between the $B^2\Sigma_u^+(v=0)$ and $X^2\Sigma_g^+(v=1)$ states corresponding to 428 nm wavelength [19] is indicated. (c) Experimentally measured spectrum of the lasing emission at 428 nm separated from the supercontinuum spectrum. (d) Far-field profile of the 428 nm laser pulse.

Figure 3. (Color online) (a) Experimentally measured polarization properties of self-seeding 428 nm laser. (b) Measured dependence of the 428 nm lasing signal on the input pump power.

Figure 4. (Color online) (a) Experimental setup of the cross-correlation measurement of 800 nm reference pulse and 428 nm lasing pulse. (b) Typical spectrum of the sum frequency signal of the 800 nm reference pulse and the 428 nm lasing pulse at a delay time around -0.5 ps. (c) Experimentally measured time-resolved sum frequency signal.

Figure 5. (Color online) Generation of self-seeding lasers at 428 nm and 391 nm by femtosecond laser filamentation in air.

**References**

[1] Rodriguez M, Bourayou R, Méjean G, Kasparian J, Yu J, Salmon E, Scholz A, Stecklum B, Eislöffel J, Laux U, Hatzes A P, Sauerbrey R, Wöste L and Wolf J P 2004 Kilometer-range nonlinear propagation of femtosecond laser pulses *Phys. Rev. E* **69** 036607

[2] Méchain G, Méjean G, Ackermann R, Rohwetter P, André Y B, Kasparian J, Prade B, Stelmaszczyk K, Yu J, Salmon E, Winn W, Schlie L A V, Mysyrowicz A, Sauerbrey R, Wöste L and Wolf J P 2005 Propagation of fs-TW laser filaments in adverse atmospheric conditions *Appl. Phys. B* **80** 785–789

[3] Méjean G, Kasparian J, Yu J, Salmon E, Frey S, Wolf J P, Skupin S, Vinçotte A, Nuter R, Champeaux S and Bergé L 2005 Multifilamentation transmission through fog *Phys. Rev. E* **72** 026611

[4] Ackermann R, Méjean G, Kasparian J, Yu J, Salmon E and Wolf J P 2006 Laser filaments generated and transmitted in highly turbulent air *Opt. Lett.* **31** 86–88

[5] Salamé R, Lascoux N, Salmon E, Kasparian J and Wolf J P 2007 Propagation of laser filaments through an extended turbulent region *Appl. Phys. Lett.* **91** 171106

[6] Kasparian J, Rodriguez M, Méjean G, Yu J, Salmon E, Wille H, Bourayou R, Frey S, André Y B, Mysyrowicz A, Sauerbrey R, Wolf J P and Wöste L 2003 White-Light Filaments for Atmospheric Analysis *Science* **301** 61-64

[7] Xu H L and Chin S L 2011 Femtosecond laser filamentation for atmospheric sensing *Sensors* **11**, 32

[8] Xu H L, Daigle J F, Luo Q and Chin S L 2006 Femtosecond laser-induced nonlinear spectroscopy for remote sensing of methane *Appl. Phys. B* **82** 655

[9]  Xu H L, Kamali Y, Marceau C, Simard P T, Liu W, Bernhardt J, Méjean G, Mathieu P, Roy G, Simard J R and Chin S L 2007 Simultaneous detection and identification of multigas pollutants using filament-induced nonlinear spectroscopy *Appl. Phys. Lett.* **90** 101106

[10] Luo Q, Xu H L, Hosseini S A, Daigle J F, Théberge F, Sharifi M and Chin S L 2006 Remote sensing of pollutants using femtosecond laser pulse fluorescence spectroscopy", *Appl. Phys. B* **82** 105–109

[11] Xu H L, Méjean G, Liu W, Kamali Y, Daigle J F, Azarm A, Simard P T, Mathieu P, Roy G, Simard J R and Chin S L 2007 Remote detection of similar biological materials using femtosecond filament-induced breakdown spectroscopy *Appl. Phys. B* **87** 151

[12] Xu H L, Liu W and Chin S L 2006 Remote time-resolved filament-induced breakdown spectroscopy of biological materials *Opt. Lett.* **31** 1541

[13] Daigle J F, Méjean G, Liu W, Théberge F, Xu H L, Kamali Y, Bernhardt J, Azarm A, Sun Q, Mathieu P, Roy G, Simard J R and Chin S L 2007 Long range trace detection in aqueous aerosol using remote filament-induced breakdown spectroscopy *Appl. Phys. B* **87** 749

[14] Daigle J F, Mathieub P, Royb G, Simardb J R and Chin S L 2007 Multi-constituents detection in contaminated aerosol clouds using remote-filament-induced breakdown spectroscopy *Opt. Commun.* **278** 147

[15] Xu H L, Bernhardt J, Mathieu P, Roy G and Chin S L 2007 Understanding the advantage of remote femtosecond laser-induced breakdown spectroscopy of metallic targets *J. Appl. Phys.* **101** 033124

[16] Xu H L, Simard P T, Kamali Y, Daigle J F, Marceau C, Bernhardt J, Dubois J, Châteauneuf M, Théberge F, Roy G and Chin S L 2012 Filament-induced breakdown remote spectroscopy in a polar environment *Laser Phys.* **22**, 1767

[17] Luo Q, Liu W and Chin S L 2003 Lasing action in air induced by ultra-fast laser filamentation *Appl. Phys. B* **76**, 337

[18] Kartashov D,Ališauskas S, Andriukaitis G, Pugžlys A, Shneider M, Zheltikov A, Chin S L and Baltuška A 2012 Free-space nitrogen gas laser driven by a femtosecond filament *Phys. Rev. A* **86** 033831

[19] Yao J, Zeng B, Xu H, Li G, Chu W, Ni J, Zhang H, Chin S L, Cheng Y and Xu Z 2011 High-brightness switchable multiwavelength remote laser in air *Phys. Rev. A* **84** 051802(R)

[20] Chu W, Zeng B, Yao J, Xu H, Ni J, Li G, Zhang H, He F, Jing C, Cheng Y and Xu Z 2012 Multiwavelength amplified harmonic emissions from carbon dioxide pumped by mid-infrared femtosecond laser pulses *Europhys. Lett.* **97**, 64004

[21] Hemmer P R, Miles R B, Polynkin P, Siebert T, Sokolov A V, Sprangle P and Scully M O 2011 Standoff spectroscopy via remote generation of a backward-propagating laser beam *Proc. Natl. Acad. Sci. U.S.A.* **108** 3130–3134

[22] Sprangle P, Peñano J, Hafizi B, Gordon D and Scully M 2011 Remotely induced atmospheric lasing *Appl. Phys. Lett.* **98** 211102

[23] Ni J, Chu W, Zhang H, Jing C, Yao J, Xu H, Zeng B, Li G, Zhang C, Chin S L, Cheng Y and Xu Z 2012 Harmonic-seeded remote laser emissions in $N_2$-Ar, $N_2$-Xe and $N_2$-Ne mixtures: a comparative study *Opt. Express* **20** 20970

[24] Yao J, Li G, Jing C, Zeng B, Chu W, Ni J, Zhang H, Xie H, Zhang C, Li H, Xu H, Chin S L, Cheng Y and Xu Z 2013 Remote creation of coherent emissions in air with two-color ultrafast laser pulses *New J. Phys.* **15** 023046

[25] Ni J, Chu W, Jing C, Zhang H, Zeng B, Yao J, Li G, Xie H, Zhang C, Xu H, Chin S L, Cheng Y and Xu Z 2013 Identification of the physical mechanism of generation of coherent $N_2^+$ emissions in air by femtosecond laser excitation *Opt. Express* **7** 8746

[26] Wille H, Rodriguez M, Kasparian J, Mondelain D, Yu J, Mysyrowicz A, Sauerbrey R, Wolf J P and Wöste L 2002 Teramobile: a mobile femtosecond-terawatt laser and detection system *Eur. Phys. J.-Appl. Phys.* **20** 183–190

[27] Liu W and Chin S L 2005 Direct measurements of the critical power of Ti:Sapphire laser pulse in air *Opt. Express* **13**, 5750

Figure 1

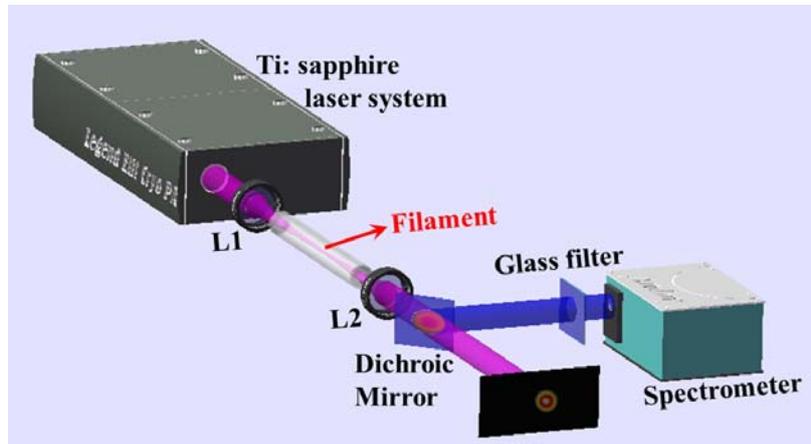

Figure 2

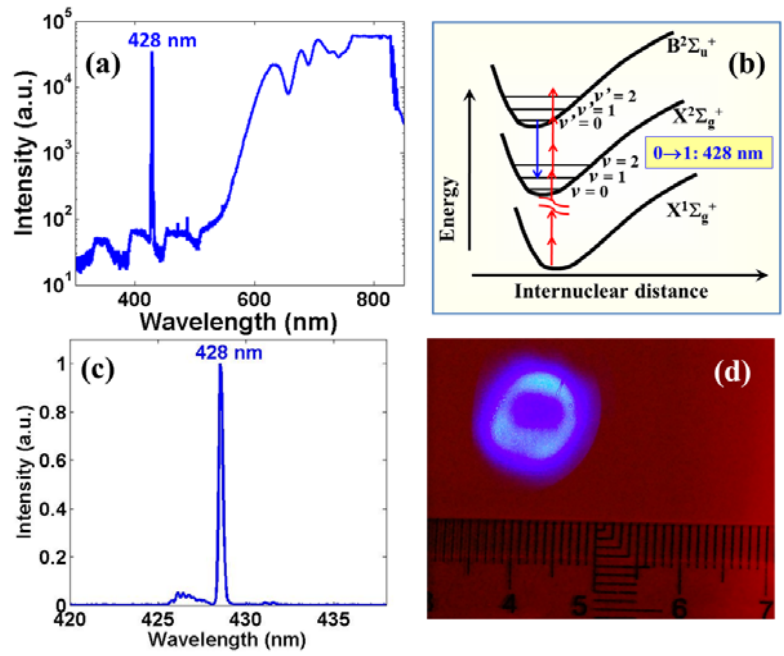

Figure 3

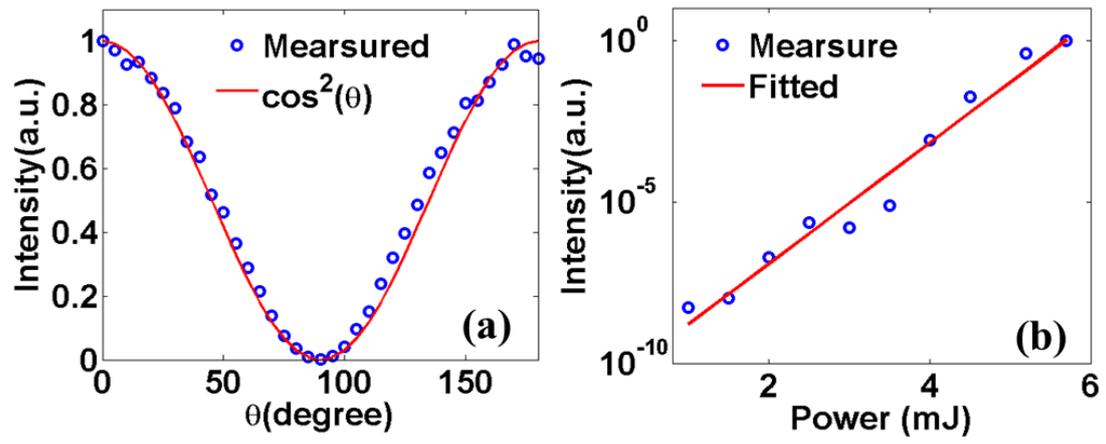

Figure 4

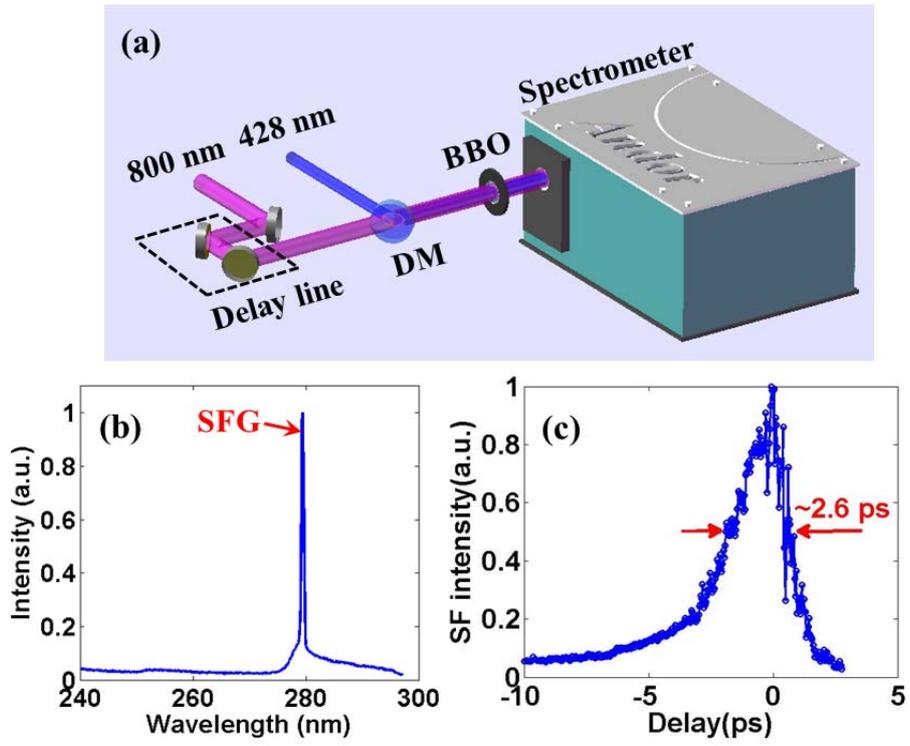

Figure 5

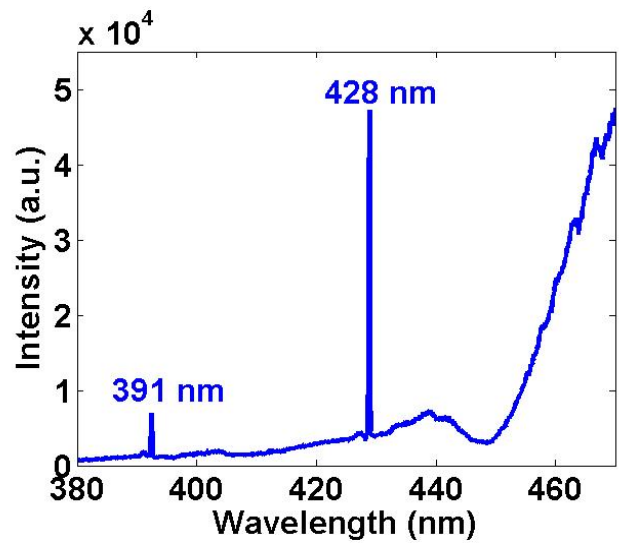